# A Primarily Survey on Energy Efficiency in Cloud and Distributed Computing Systems

Nikzad Babaii Rizvandi[1,2],  Albert Y. Zomaya [1]

[1] Center for Distributed and High Performance Computing, School of Information Technologies, University of Sydney, Sydney, Australia

[2] National ICT Australia (NICTA), Australian Technology Park, Sydney, Australia

nikzad@it.usyd.edu.au
albert.zomaya@sydney.edu.au

## 1 Introduction

Research on low power systems has received a great amount of attention in recent years since the sustainability of current technologies and practices has become a serious issue. A few example systems where lowering power usage is critical are:

- *Wireless sensors:* several sensors extract data from the environment concurrently, transmit these data to a processing unit and receive processed data accompanied by appropriate commands from the processing unit [1-5]. The sensors and their receiver/transmitter are generally powered by battery and/or solar cells.
- *Satellite circuits:* Satellites typically involve massive number of complex circuits that must work in low power. These circuits are supplied by solar cells, the only available power supply in satellites.
- *Robots and surveillance devices:* these devices are heavily used in army, mine extraction and in difficult or unsafe environments for humans.
- *Cell phones and laptops:* these devices are powered by batteries which are expected to work for a long time.

In the meantime, stiff increases in energy price and the environmental impact of carbon dioxide emissions associated with energy generation and transportation have forced the issue of reducing energy consumption to be extended to a broader range of system including High Performance Computing Systems (HPCS).

Various issues such as resource management in both software and hardware levels must be addressed to reduce energy consumption in HPCS. An important issue in hardware resource management is how to reduce power usage in processors. In the recent past, many hardware-based approaches have been made to efficiently reduce energy consumption, particularly for processors. Dynamic voltage-frequency scaling (DVFS) is perhaps the most appealing method incorporated into many recent processors. Energy savings with this method is based on the fact that the power consumption in CMOS circuits has direct relation with frequency and the square of voltage supply. In this case, the execution time and power consumption can be controlled by switching between processor's frequencies and voltages. Although this approach was initially designed for single processor task scheduling [6], it has recently received much attention in multiprocessor systems as well [7, 8].

DVFS technique and task scheduling can be combined in two ways: (1) schedule generation, and (2) slack reclamation. In the schedule generation, tasks graph are (re)scheduled on DVFS-enabled processors in a global cost function including both energy saving and makespan to meet both energy and time constraints at the same time [9-12]. In slack reclamation, which works as post processing procedure on the output of scheduling algorithms, DVFS technique is used to minimize the energy consumption of tasks in a schedule generated by a separate scheduler. The existing methods based on DVFS technique, however, have two major shortcomings: (1) most of them focus on schedule generation and do not adequately take the slack reclamation approaches into account to save more energy, and (2) the existing slack reclamation methods use only one frequency for each task among all discrete set of processor's frequencies. Using one frequency usually results in uncovered slack time where processor and other devices only waste energy.

## 2 Energy Efficiency in HPCS

Many of electronic systems in our life such as satellite systems, cell-phones, game instruments and so on are using rechargeable batteries as their power supplies. Although the battery capacity has been grown significantly in recent years (the battery capacity increases 5% per year), battery life is still the major drawback for most of electronic systems. In addition to power-aware battery-based systems, the issue of energy consumption has recently attracted a great amount of attention in high performance computing systems (HPCS). Energy consumption issue in such systems can be classified into three groups: (1) system-level resource allocation, (2) service-level energy-load distribution, and (3) task scheduling level (Figure 2-1).

In the system-level, the problem is how to distribute computational resources (e.g. CPU, network, memory and I/O) between large scale data storages and processing centers (such as supercomputers and data centers). Fairly distribute resources among applications (or services) not only requires to obtain individual adaptation among resources but also needs to understand the interaction between individual resources when they work as a system. Therefore, the big challenge here is to find both the relationship among system resources and their trade-off, which may cause an optimal balance between performance, QoS and energy consumption [13]. Among different technologies in system-level for managing resources between workloads, virtualization becomes a key technology in data centers. Virtualization allows the computational resources to be shared between different workloads. Many of incoming workloads to data centers are medium size workloads which often require a small fraction of the computational resources. The servers typically spend around 70% of their maximum power consumption even in low utilization. With virtualization, such workloads can be run within a virtual machine (VM) causing significant saving in overall energy usage. The associated VMs may require fewer amounts of resources and therefore they can be run on a single hardware unit. It is obvious that less hardware is used in overall, less energy is wasted for both working on and cooling of the servers.

In the service-level, energy reduction by load balancing, scheduling and mapping workloads is concerned. The main challenge is to utilize appropriate algorithms to both multiplex/demultiplex workloads in order to save energy and make a trade-off between performance and service cost reduction because of energy savings. Also, to avoid hotspot in data centers due to high-loaded nodes, services can be moved from nodes with high-load and high temperature to nodes with smaller load and lower temperature. Generally, this movement of services should happen when the destination nodes can operate the services in an energy efficient way [13].

In site-level/hardware level, the focus of this article, the operating system (OS) and hardware configuration such as dynamic power management, micro-architecture techniques and dynamic voltage scaling are used to decrease power. Here, the typical question could be: *"What is the suitable OS/hardware configuration to process tasks in the shortest possible time and with minimum energy?"*

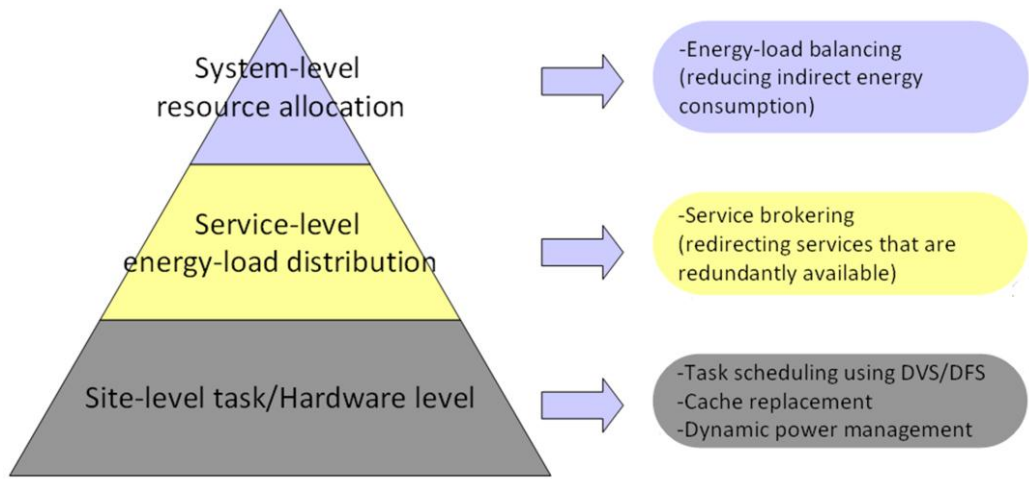

Figure 2-1. Energy consumption levels in HPCS

# 3 Dynamic Power Management (DPM)

Dynamic power management (DPM) is an operating system level mechanism to dynamically configure the hardware. This mechanism tries to use minimum amount of active components for the requested services and performance [14]. Generally, computing systems and their components are utilized non-uniformly by workloads during operation time. Therefore by dividing system operation time into different working states, the next state of workload can be predicted regarding to the previous states of workload with combination of the current state.

DPM is a set of computational techniques to save power by selectively turning on and off of system components or moving them low-power states in idle time. Typically, there is a power manager (PM) to control the components working states. PM makes decision based on observing the workload behaves during runtime and assuming some assumption about the workload. PM then follows some policies to manage power. For example, one simple policy which is already used in Laptops is turning off the system after spending pre-defined seconds of time in idle. In literation, several works on DPM-based task scheduling can be found. In [15], authors present an online greedy scheduling algorithm for independent tasks. The algorithms attempts to reduce energy consumption by ordering task execution so that devices can have continuous long idle periods to be shut down. Moreover, reducing the number of device on/off shrinks transition delays between device operating states. Unlike most real-time DPM techniques which are CPU-enteric, authors in [16] focus on I/O devices by proposing an offline branch-and-bound algorithm regard to finding optimal energy schedule for a given task set and meeting all real-time deadlines. As operating system (OS) observes the relationship between processes and hardware devices, in [17], an approach to reduce energy consumption of I/O devices in interactive systems is proposed. After estimating the utilization of a device from each running process on the system by the OS, a device leads to a low-power state if it is not running any process. Also, it is proved that even when no timing dependency is considered solving optimal low-power task scheduling for DPM on multiple devices is categorized as an NP hard problem.

### 3.1 Simple shutdown

This is important to note that clock gating, a power-saving technique utilized in many synchronous circuits, does not decrease the loss of power [14]. First, power loss still exists in clock circuit when there is local clock gating or the clock generator is active. Second, leakage currents in components always waste power even all clocks are off. Therefore, clock gating may be inefficient to save power in battery-based devices. A simple but useful policy for battery-based devices is simply turning off them in the idle time. The main benefit of this policy is that it can be used in almost all electronic systems both digital and analog, sensors and transducers. However, the major drawback is the time of activating the device (as the components must be initialized) which typically is more than the clock gating.

### 3.2 OnNow

Microsoft OnNow is a power management tool in operating system level which supports OSPM (Operating System-directed configuration and Power Management) in personal computers with capability to move to different power states. This method uses hardware capability of computers to move to sleep mode instead of turning off [14]. The general policy in OSPM is that operating system automatically sends computer to sleep mode when computer is in idle for some time or when user press a button on the front of computer to indicate that the session has been done. In sleep mode, processor and peripherals do not work but some components are still working for capturing events. If an event happens in either software or hardware, the system wakes up and goes to active state.

With OnNow, operating system can control the global power-state transitions. Depending on the global power policy, the system may move to sleep depending on the reports coming from hardware. Also, the current running applications can be configured to wake up the system. As each component of the system has its own power states, it can go to sleep independent from other components even when some parts of system are working. Therefore, another policy of system is to adapt special requirements of each application with both system capabilities and operating system information in order to save energy without harmful influence on the work that user is doing.

**3.3 Advanced Configuration and Power Interface (ACPI)**

Advanced Configuration and Power Interface (ACPI), supported by Intel, Microsoft and Toshiba, is a standard to simplify OSDM by introducing standard interface to control system recourses [14]. This standard, however, does not provide a unique and proper procedure to control all systems efficiently. For example, ACPI-related policies are different between desktop computers and laptops. So it is necessary to apply different policies on a system and examine which combination of policies can help the system work energy efficient. A study to find the best policies for desktop and laptop has been published in [18]. Among existing operating systems, Microsoft Windows 98 is the only one fully supports ACPI, while Windows XP/Vista/200, Linux and desktop version of SunOS partially support ACPI [14].

# 4 Micro-Architecture Techniques

The increase of circuits' temperature due to the sharp increase of the number of transistors per cm2 has led the circuit designers to enhance and modify the architecture and design of microprocessors (such as on-chip system integrations, multi-clock frequency implementation, multiple voltages). As in CMOS circuits, there is a direct relation between the clock frequency and power consumption and consequently circuit temperature, the issue of power consumption in a circuit can easily limit the maximum clock frequency. In another word, in order to face this problem, power/performance and power/cost optimization must be considered for designing new microprocessor. Two major micro-architecture techniques to power reduction are Adaptive architecture and cache management [14] as follows.

**4.1 Adaptive architecture**

Since there is strong dependency between the performance of DPM policies and workload statistics, static prediction of workload becomes ineffective when the workload is either unknown or non-stationary. Therefore, DPM requires some form of adaptation. The idea of adaptation is based on predicting of workload timer duration or sometimes using several timeout for non-stationary workloads. Krishnan et al in [19] use a set of timeout values where each timeout is associated with an index to show how well this timeout value is. Then the best timeout which performs best among the set of available ones is chosen as the adaptation policy. Another work was done by Helmbold et al in [20] which keeps a list of candidate timeouts and assigns a weight to each timeout. The actual timeout is the weighted average of all candidates. To find more information about prediction policies refer to Langen and Juurlink work in [21].

**4.2 Cache management**

Memory is one of the power-hungry components in most of computer systems. Based on a report from IBM eServer machine [22] memory spends about 40% of the total energy which ranks it as the second power consumer in computing systems after processor. In addition to computing systems, memory is still one of the biggest power consumers in PDAs and laptops. The reduction of memory power is important as in most of computing systems memory works and consumes power continuously.

Power-Aware Virtual Memory (PAVM), proposed by Huang et al. in [23], is a recent powerful scheme to reduce memory power consumption by allocating active pages to the same memory. This scheme is used for the main memory and at operating system level. The idea behind this schema is that it turns on the active ranks of p and turns off inactive ranks while p is being scheduled. A rank is defined as an active rank of process p if and only if at least one page from the rank is mapped into the address space of p. The main drawback of this schema is the amount of buffer cache it uses. Buffer caches are used as a bridge between hard disk and main memory. Increasing total memory size results in increasing amount of buffer cache and therefore occupying bigger part of the memory. To find more information about different cache management techniques based on PAVM scheme refer to [22-24].

# 5 Dynamic Voltage-Frequency Scaling (DVFS)

Dynamic voltage-frequency scaling is a modern technique in computer architecture to reduce energy consumption of microprocessors or control amount of generated heat by circuits. This technique is commonly utilized in battery-based devices such as laptops and cell phones where decreasing the energy usage of battery is necessary. In addition, DVFS is used in high-computing nodes not only to decrease the power of the nodes but also to save more energy to cool down the nodes' places. An approximation model shows that the dynamic power in CMOS circuits is a linear function of both switching frequency and voltage square as: $C.f.V^2$, where C is the effective switching capacity per clock cycle. Therefore, a workload (or task) can save more energy when it is executed in lower voltage and frequency. In general, a computing node executes several tasks with inter-task relationships (e.g., precedence constraints) simultaneously. These inter-task relationships typically incur slack time (idle time) between tasks where can be used by DVFS to reduce energy usage. Specifically, the slack time associated with a task is utilized to execute the task in a lower voltage-frequency; this in turn results in energy reduction.

In recent years, there has been a significant amount of work on low-power task scheduling using DVFS technique, especially in real-time embedded systems and HPCS. The main idea of most existing algorithms is that the processor slack time should be filled by switching the processor operating frequency to lowest possible frequency; this changes the DVFS problem to estimating of the processors' slack time and the tasks' timing information (e.g. task deadline, task release time and task execution time). Based on the tasks' timing information availability, energy-aware task scheduling in embedded systems is categorized into two groups: real-time scheduling and non-real-time scheduling.

In real-time scheduling, task release time (i.e., arrival time) and total number of CPU cycles needed to complete each task (i.e., task execution time) are unknown a priori. There are many studies to apply DVFS in real-time scheduling scenarios. In [25] and [26] low-power scheduling of multi tasks while meeting given hard timing constraints in the operating system (OS) has been studied. The main assumption in these coarse-grained DVFS approaches – which is practically difficult to achieve– is that the total number of CPU cycles required to execute each task is fixed and available as a priori. In [27], an energy aware scheduling algorithm is formulated into a mathematical programming formulation of the scheduling and voltage/frequency assignment problem. To speed up the algorithm, a schedule table at design time is constructed to provide multiple scheduling options for each task; these multiple scheduling may use complex algorithms to build the schedule table at design time.

A stochastic allocation, non-linear integer programming based scheduling algorithm is proposed in [28] for multiprocessor platforms. In addition to exact analytic formulation of the stochastic objective function based on the task graph analysis for scheduling, the authors also extend the timetable constraint for conditional activities to improve their stochastic resource allocation. In [29], probabilistic distribution of all tasks' execution time is used to better partition the workload and consequently reduce more energy consumption. Then a polynomial-time heuristic method is proposed to convert the problem of scheduling, which is NP-hard, into a probability-based load balancing problem; this problem then is solved with worst-fit decreasing bin-packing heuristic. Authors in [30] consider uncertainty in execution times of tasks and model each varied execution time as a random variable. Then two algorithms are proposed for both uniprocessor and multiprocessor using probabilistic approach; these algorithms are designed to both minimize total energy consumption of the system and satisfy timing constraints with a guaranteed confidence probability. However, their algorithms for multiprocessor platforms suffer from exponential complexity.

In energy harvesting systems – embedded systems working in environment without access to energy power–, low-power scheduling using DVFS increases their lifetime. In such systems, tasks are periodic and generally simple; also their schedulers try to find an optimal trade-off between energy consumption and deadline miss rate of tasks – which means time constraint are not hard in such systems. In [31], Adaptive Scheduling DVFS (AS-DVFS) is proposed which simplifies the original scheduling problem by decoupling/separating timing and energy constraints; then, AS-DVFS adjusts the processor/processors operating frequency (1) based on workload information, and (2) under the timing and energy constraints towards achieving the whole system energy efficiency. Scheduling algorithm in [32] reduce energy consumption by (1) assigning tasks to the processing element with lower operating frequency, and (2) migrating these tasks among processors. Authors in [33] propose UTilization Based (UTB) algorithm by combining energy harvesting awareness, DVFS, and task slack management to reduce energy consumption of periodic tasks on multiprocessor systems. The proposed a low-complexity task scheduling algorithm is based on the concept of task CPU utilization – defined as the worst-case task execution time divided by its period.

In [34], an intra-task voltage scheduling algorithm based on a static timing analysis of an application is proposed; at first, a given task is partitioned into several segments, and then appropriate supply voltage (resulting from static timing analysis of previous segments and based on worse-case execution time of the task) is assigned for each segment. The authors claim that this scheduling algorithm has a high energy reduction ratio by fully exploiting all the slack times, and choosing suitable voltage/frequency to fill these slack times as much as possible. A software feedback loop based method is proposed in [35] where for each time slot a deadline is provided. Then, processor operating frequency of current slot is calculated based on (1) the slack time generated in the previous slot, and (2) the worst-case execution time of the current slot. The main assumption in both mentioned DVFS methods is that the worse-case execution time of each segment/slot of a task is known which is difficult to reach in many applications; for instance, in MPEG decoding it is too hard to precisely estimate/calculate the worse-case execution time of each frame. Although calculating a single unique worse-case execution time for all frames in a specific video can be an option – and also easy–, it cannot significantly use the potential of DVFS-based techniques to reduce energy consumption. Authors in [36] focus on energy efficient MPEG decoding by proposing an effective DVFS algorithm based on future workload prediction; frame-based history is used to predict computational workload of an incoming frame. This prediction gives this opportunity to pick suitable voltage/frequency of the processor, and provide the exact amount of computing power to decode the frame. Therefore, decoding time of a frame is divided into two parts: frame-dependent part and frame-independent part. The statistical variation in dependent-part of a frame is compensated by its independent-part; this allows a considerable energy saving with less degradation of QoS.

While in non-real-time scheduling, timing information of tasks is available in compile time. Having this information allows schedulers (such as list scheduler) to be developed by maximizing processor utilization with meeting all deadlines [6, 9]. This type of scheduling involves most of the large-scale computational problems such as bioinformatics, chemistry and object recognition in machine vision applications [37]. In [38], tasks are first scheduled using list-scheduling algorithm where the mutual of the slack time is used as the tasks' priorities. Then, these slack times are distributed among tasks on each critical path; thus processors can move to lower frequency/voltage and consume less energy. In [39] scheduled tasks are re-scheduled by an extended list-scheduling algorithm. At each time step, the algorithm first calculates energy saving of a task when it is scheduled at the current step and the next step. The difference between the energy of these two steps are represented as the task's energy saving. Then a task with a higher energy saving and lower slack time gets higher priority to be scheduled. A two-phase solution is proposed in [40] where a version of early-deadline-first scheduling algorithm is utilized in the first phase for assigning a task to a best-fit processor regard to the task ready time and the processor free time; in the second phase, the proper voltage/frequency of the processor to run each task is then is solved by an Integer Linear Programming (ILP).

There are many algorithms in the literature for energy-efficient real-time and non-real-time scheduling embedded systems – where these algorithms are suitable for platforms with small number of processors and mostly assume the tasks (periodic or aperiodic) are independent; however, a few algorithms have been developed for reduction of power consumption in HPCS using DVFS and DPM. In [41], the authors present a theoretical framework for energy-efficient e-business datacenters and introduce an automatic power and performance management methodology; their methodology uses mathematically-rigorous optimization technique to optimize performance/watt regard to meeting performance constraints and minimizing energy consumption. There are a few optimization techniques for online scheduling algorithms generally by deploying popular heuristic algorithms like MET, Min-Min, Max-Min, OLB, or fast greedy [42, 43]. Moreover, in [42, 44, 45] meta-heuristic algorithms such as Simulated Annealing and Tabu have been proposed.

Authors in [46] study the effect of virtualization for power-aware consolidation with proposing a dynamic configuration technique/algorithm to effectively optimize the power in virtualized server clusters and dynamically manage it. In the case of consolidation of multiple services/applications on such virtualized clusters, a dynamic configuration was developed based on a mixed integer programming (MIP) formulation [47]. This power-efficient approach also takes into account the cost of frequently on/off the servers. However, this approach is too slow to make a proper decision for an online scheduler. In [48], Virtual Machine (VM) is used for executing HPC applications to reduce energy consumption of virtualized datacenters by supporting VM migration and VM placement optimization with less human interaction; also overhead of virtualization is taken into account. The usage of different datacenters with distributed locations has been presented in [49] in order to distribute workload among these datacenters regard to minimizing power consumption and guarantee SLA. Another dynamic job scheduling policy is proposed in [50] for power-aware resource allocation in a virtualized datacenter. In addition to considering into account the overhead of virtualization, the technique tries to gather workloads from separate machines into a smaller number of nodes, turns off more servers and thus reduces the overall datacenter power consumption.

In [51], Just-in-time DVFS technique was presented to fill slack time in MPI programs. A system called Jitter was utilized to reduce the frequency on nodes with more slack times and fewer computations. The goal of Jitter was to be sure that they came just in time without increasing overall execution time. Ge et al. in [7] applied DVS technique on processors that did not work in peak performance during execution of a parallel application. The best processor frequency of each task was selected by analyzing computation and communication power profiles collected before execution. A method to reduce power consumption was presented in [10] by adaptively activating and deactivating hardware resources, especially memory for intensive HPC applications. Cache missing in accessing the main memory also plays a great role in adjusting and triggering processors slack times.

Lee and Zomaya in [10] presented a DVFS-based algorithm to minimize both completion time and energy consumption of precedence-constrained parallel jobs on HPC systems. This method tried to minimize a summation of two cost functions: completion time and energy.

Therefore, the final result was a trade-off between the quality of scheduling and energy consumption. In [52, 53] multiple frequency selection algorithm is presented. The idea is that instead of execution of a task with only one frequency, the task is executed with a linear combination of all frequencies in the DVFS-enabled processor. This results in covering more slack between tasks and therefore more energy saving. The authors in [54] study this multiple selection mathematically. They realized that a task can reach to its minimum energy consumption with it is executed with a linear combination of at most two frequencies (not all frequencies). When a simplified version of DVFS is used (i.e., frequency has proportional relation with voltage) these two frequencies are adjacent. Ding et al. in [55] introduced the concept of energy scalability in formal terms. In addition to study energy efficiency/iso-efficiency concept, they extended an analytical model to study tradeoffs between performance and energy saving in HPCS. In [56], the slack times in real-time applications were classified into static, work and shared lack groups for multiple dependent tasks on multiple DVFS-enabled processors. Then a dynamic dependency aware task scheduling was proposed to adjust voltage/frequency of each processor regarding tasks' real time deadlines. Hotta [57] presented a profiled-based power-performance optimization method utilizing DVFS in HPCS. Here, the execution of a program was divided into several regions. In trial steps, profile information of each region (including power and execution profiles) was extracted and then utilized to find its best combination of processors' voltages and frequencies.

Authors in [58] present their power-aware scheduling algorithms for bag of-tasks applications with deadline constraints on DVS-enabled cluster systems; this scheduling algorithms tries to both minimize power consumption and meet the deadlines specified by application users. In [59], an upper limit for system energy usage was chosen externally. Then a combination of performance modeling and performance prediction was used to reduce execution times with respect to their predefined energy usage upper limit. After creating models for both execution time and energy consumption, key parameters of models were estimated by executing a program for a small number of times and then regressing the estimated parameters. Here, for better estimation of parameters, the following steps were iterated until a proper schedule is achieved: (1) using models to predict each possible scheduling of tasks, (2) executing the program a few times with the best predicted schedule and (3) updating estimated key parameters. Rountree et al in [60] proposed an energy-aware schedule generation algorithm for DVFS-enabled processors where a combination of all processor frequencies is involved into an overall linear programming optimization. An energy reduction algorithm was proposed by Kimura et al in [8] for power scalable high performance cluster supporting DVFS. In a simplified version of this algorithm, the suitable frequency among discrete set of processor's frequencies are chosen for each task regarding to each task's slack time.

### 5.1 DVFS and DPM combination techniques

It is worth noting that there are several works in literature to reduce power consumption by combining both DVFS and DPM. A Markovian decision processes based DPM model is presented by the authors in [61] which is a uniform modelling framework for both DVS and DPM. Another stochastic approach is proposed in [62] where the authors combine DVS and renewal theory based DPM approach. Despite the effectiveness, these two stochastic approaches cannot handle tasks with hard deadline constraints or dependency; the hard deadline constraint is addressed in [63]

## References

[1] N. Kamyabpour and D. B. Hoang, "A hierarchy energy driven architecture for wireless sensor networks," presented at the 24th IEEE International Conference on Advanced Information Networking and Applications (AINA-2010), Perth, Australia, 2010.

[2] N. Kamyabpour and D. B. Hoang, "A Task Based Sensor-Centeric Model for overall Energy Consumption," *CoRR,* 2012.

[3] K. Almiani, S. Selvakennedy, and A. Viglas, "RMC: An Energy-Aware Cross-Layer Data-Gathering Protocol for Wireless Sensor Networks," presented at the 22nd International Conference on Advanced Information Networking and Applications (AINA), GinoWan, Okinawa, Japan, 2008.

[4] K. Almiani, A. Viglas, and L. Libman, "Energy-efficient data gathering with tour length-constrained mobile elements in wireless sensor networks," presented at the The 35th Annual IEEE Conference on Local Computer Networks (LCN), Denver, Colorado, USA, 2010.

[5] N. Kamyabpour and D. B. Hoang, "A study on Modeling of Dependency between Configuration Parameters and Overall Energy Consumption in Wireless Sensor Network (WSN)," *CoRR,* 2011.

[6] J. Zhuo and C. Chakrabarti, "Energy-efficient dynamic task scheduling algorithms for DVS systems," *ACM Trans. Embed. Comput. Syst.,* vol. 7, pp. 1-25, 2008.

[7] R. Ge, X. Feng, and K. W. Cameron, "Performance-constrained Distributed DVS Scheduling for Scientific Applications on Power-aware Clusters,"

presented at the Proceedings of the 2005 ACM/IEEE conference on Supercomputing, 2005.

[8] H. Kimura, M. Sato, Y. Hotta, T. Boku, and D. Takahashi, "Emprical study on Reducing Energy of Parallel Programs using Slack Reclamation by DVFS in a Power-scalable High Performance Cluster," in *Cluster Computing, 2006 IEEE International Conference on*, 2006, pp. 1-10.

[9] R. Xiaojun, Q. Xiao, Z. Ziliang, K. Bellam, and M. Nijim, "An Energy-Efficient Scheduling Algorithm Using Dynamic Voltage Scaling for Parallel Applications on Clusters," in *Computer Communications and Networks, 2007. ICCCN 2007. Proceedings of 16th International Conference on*, 2007, pp. 735-740.

[10] Y. C. Lee and A. Y. Zomaya, "Minimizing Energy Consumption for Precedence-Constrained Applications Using Dynamic Voltage Scaling," presented at the Proceedings of the 2009 9th IEEE/ACM International Symposium on Cluster Computing and the Grid (CCGrid), 2009.

[11] Y. Kessaci, N. Melab, and E.-G. Talbi, "A pareto-based GA for scheduling HPC applications on distributed cloud infrastructures " presented at the International Conference on High Performance Computing and Simulation (HPCS), Istanbul 2011.

[12] Y. Kessaci, M. Mezmaz, N. Melab, E.-G. Talbi, and D. Tuyttens, "Parallel Evolutionary Algorithms for Energy Aware Scheduling " in *Intelligent Decision Systems in Large-Scale Distributed Environments*, ed: Springer Vlg, 2011, pp. 75-100.

[13] A. Berl, E. Gelenbe, M. d. Girolamo, G. Giuliani, H. d. Meer, M. Q. Dang, and K. Pentikousis, "Energy-Efficient Cloud Computing," *The Computer Journal,* 2009.

[14] L. Benini and G. d. Micheli, *Dynamic Power Management: Design Techniques and CAD Tools*. Norwell, MA, USA: Kluwer Academic Publishers, 1998.

[15] Y.-H. Lu, L. Benini, and G. D. Micheli, "Low-power task scheduling for multiple devices," presented at the Proceedings of the eighth international workshop on Hardware/software codesign, San Diego, California, United States, 2000.

[16] V. Swaminathan and K. Chakrabarty, "Pruning-based, energy-optimal, deterministic I/O device scheduling for hard real-time systems," *ACM Trans. Embed. Comput. Syst.,* vol. 4, pp. 141-167, 2005.

[17] Y.-H. Lu, L. Benini, and G. D. Micheli, "Power-aware operating systems for interactive systems," *IEEE Trans. Very Large Scale Integr. Syst.,* vol. 10, pp. 119-134, 2002.

[18] E. Y. C. Y. Lu, T. Simunic, L. Benini and G. De Micheli "Quantitative Comparison of Power Management Algorithms," *Proceedings of Design Automation and Test in Europe (DATE),* March 2000.

[19] P. L. P. Krishnan, J. Vitter, "Adaptive Disk Spindown Via Optimal Rent-to-buy in Probabilistic Environments," *International Conference on Machine Learning,* pp. 322-330, July 1995.

[20] D. L. D. Helmbold, E. Sherrod, "Dynamic Disk Spin-down Technique for Mobile Computing," *Conference on Mobile Computing,* pp. 130-142, Nov.1996.

[21] P. d. Langen and B. Juurlink, "Trade-Offs Between Voltage Scaling and Processor Shutdown for Low-Energy Embedded Multiprocessors," presented at the Embedded Computer Systems: Architectures, Modeling, and Simulation, 2007.

[22] E. S. Min Lee, Joonwon Lee, and Jin-soo Kim, "PABC: Power-Aware Buffer Cache Management for Low Power Consumption," *IEEE Transactions on Computers,* vol. 56, April 2007.

[23] C. L. H. Huang, T. Keller, and K.G. Shin, "Memory Traffic Reshaping for Energy-Efficient Memory," *Int'l Symp. Low Power Electronics and Design (ISLPED '05),* pp. 393-398, Aug 2005.

[24] P. P. H. Huang, and K.G. Shin, "Design and Implementation of Power-Aware Virtual Memory," *Proc. USENIX Ann. Technical Conf.,* pp. 57-70, 2003.

[25] F. Yao, A. Demers, and S. Shenker, "A scheduling model for reduced CPU energy," presented at the Proceedings of the 36th Annual Symposium on Foundations of Computer Science, 1995.

[26] T. Ishihara and H. Yasuura, "Voltage scheduling problem for dynamically variable voltage processors," presented at the Proceedings of the 1998 international symposium on Low power electronics and design, Monterey, California, United States, 1998.

[27] J. Cong and K. Gururaj, "Energy Efficient Multiprocessor Task Scheduling under Input-dependent Variation," presented at the Design, Automation & Test in Europe Conference & Exhibition (DATE '09), Dresden, Germany, 2010.

[28] M. Lombardi and M. Milano, "Stochastic allocation and scheduling for conditional task graphs in MPSoCs," presented at the Proceedings of the 12th international conference on Principles and Practice of Constraint Programming, Nantes, France, 2006.

[29] C. Xian, Y.-H. Lu, and Z. Li, "Energy-aware scheduling for real-time multiprocessor systems with uncertain task execution time," presented at the Proceedings of the 44th annual Design Automation Conference, San Diego, California, 2007.

[30] M. Qiu, C. Xue, Z. Shao, and E. H.-M. Sha, "Energy minimization with soft real-time and DVS for uniprocessor and multiprocessor embedded systems," presented at the Proceedings of the conference on Design, automation and test in Europe, Nice, France, 2007.

[31] S. Liu, Q. Wu, and Q. Qiu, "An adaptive scheduling and voltage/frequency selection algorithm for real-time energy harvesting systems," presented at the Proceedings of the 46th Annual Design Automation Conference, San Francisco, California, 2009.

[32] T. Wei, Y. Guo, X. Chen, and S.Hu, "Adaptive Task Allocation for Multiprocessor SoCs in Energy Harvesting Systems," presented at the 11th

[33] J. Lu and Q. Qiu, "Scheduling and mapping of periodic tasks on multi-core embedded systems with energy harvesting," presented at the Second International Green Computing Conference (IGCC 2011), Orlando, Florida, 2011.

[34] D. Shin, J. Kim, and S. Lee, "Low-energy intra-task voltage scheduling using static timing analysis," presented at the Proceedings of the 38th annual Design Automation Conference, Las Vegas, Nevada, United States, 2001.

[35] S. Lee and T. Sakurai, "Run-time power control scheme using software feedback loop for low-power real-time application," presented at the Proceedings of the 2000 Asia and South Pacific Design Automation Conference, Yokohama, Japan, 2000.

[36] K. Choi, W-C. Cheng, and M. Pedram, "Frame-based dynamic voltage and frequency scaling for an MPEG player," *Journal of Low Power Electronics, American Scientific Publishers,* vol. 1, pp. 27-43, April 2005.

[37] J. M. Geusebroek and F. J. Seinstra, "Object Recognition by a Robot Dog Connected to aWide-Area Grid System," in *Multimedia and Expo, 2005. ICME 2005. IEEE International Conference on*, 2005, pp. 1565-1566.

[38] J. Luo and N. K. Jha, "Static and Dynamic Variable Voltage Scheduling Algorithms for Real-Time Heterogeneous Distributed Embedded Systems," presented at the Proceedings of the 2002 Asia and South Pacific Design Automation Conference, 2002.

[39] F. Gruian and K. Kuchcinski, "LEneS: task scheduling for low-energy systems using variable supply voltage processors," presented at the Proceedings of the 2001 Asia and South Pacific Design Automation Conference, Yokohama, Japan, 2001.

[40] Y. Zhang, X. S. Hu, and D. Z. Chen, "Task scheduling and voltage selection for energy minimization," presented at the Proceedings of the 39th annual Design Automation Conference, New Orleans, Louisiana, USA, 2002.

[41] B. Khargharia, S. Hariri, and M. S. Yousif, "Autonomic power and performance management for computing systems," *Cluster Computing,* vol. 11, pp. 167-181, 2008.

[42] T. D. Braun, H. J. Siegel, N. Beck, L. L. B\, \#246, l\, ni, M. Maheswaran, A. I. Reuther, J. P. Robertson, M. D. Theys, B. Yao, D. Hensgen, and R. F. Freund, "A comparison of eleven static heuristics for mapping a class of independent tasks onto heterogeneous distributed computing systems," *J. Parallel Distrib. Comput.,* vol. 61, pp. 810-837, 2001.

[43] R. Armstrong, D. Hensgen, and T. Kidd, "The Relative Performance of Various Mapping Algorithms is Independent of Sizable Variances in Run-time Predictions," presented at the Proceedings of the Seventh Heterogeneous Computing Workshop, 1998.

[44] A. Abraham, R. Buyya, and B. Nath, "Nature's Heuristics for Scheduling Jobs on Computational Grids," presented at the 8th International Conference on Advanced Computing and Communications, India, 2000.

[45] M. Mika, G. Walig\, \#243, ra, and J. Weglarz, "A metaheuristic approach to scheduling workflow jobs on a Grid," in *Grid resource management*, N. Jarek, M. S. Jennifer, and W. Jan, Eds., ed: Kluwer Academic Publishers, 2004, pp. 295-318.

[46] V. Petrucci, O. Loques, B. Niteroi, and D. Mossé, "Dynamic configuration support for power-aware virtualized server clusters," presented at the 21th Conference on Real-Time Systems, Dublin, Ireland, 2009.

[47] V. Petrucci, O. Loques, and D. Moss´e, "A Dynamic Configuration Model for Power-efficient Virtualized Server Clusters," presented at the 11th Brazillian Workshop on Real-Time and Embedded Systems (WTR), Recife, Brazil, 2009.

[48] L. Liu, H. Wang, X. Liu, X. Jin, W. B. He, Q. B. Wang, and Y. Chen, "GreenCloud: a new architecture for green data center," presented at the Proceedings of the 6th international conference industry session on Autonomic computing and communications industry session, Barcelona, Spain, 2009.

[49] K. Le, R. Bianchini, M. Martonosi, and T. Nguyen, "Cost-and Energy-Aware Load Distribution Across Data Centers," presented at the Workshop on Power Aware Computing and Systems (HotPower'09), Big Sky, MT, USA, 2009.

[50] I. Goiri, F. Julia, R. Nou, J. L. Berral, J. Guitart, and J. Torres, "Energy-Aware Scheduling in Virtualized Datacenters," presented at the Proceedings of the 2010 IEEE International Conference on Cluster Computing, 2010.

[51] N. Kappiah, V. W. Freeh, and D. K. Lowenthal, "Just In Time Dynamic Voltage Scaling: Exploiting Inter-Node Slack to Save Energy in MPI Programs," presented at the Proceedings of the 2005 ACM/IEEE conference on Supercomputing, 2005.

[52] N. B. Rizvandi, J. Taheri, A. Y. Zomaya, and Y. C. Lee, "Linear Combinations of DVFS-enabled Processor Frequencies to Modify the Energy-Aware Scheduling Algorithms," presented at the Proceedings of the 2010 10th IEEE/ACM International Symposium on Cluster, Cloud and Grid Computing (CCGrid), Melbourne, Australia, May 17-20, 2010.

[53] N. B. Rizvandi, A. Y.Zomaya, Y. C. Lee, A. J. Boloori, and J. Taheri, "Multiple Frequency Selection in DVFS-Enabled Processors to Minimize Energy Consumption," in *arXiv:1203.5160v1*, ed, 2012.

[54] N. B. Rizvandi, J. Taheri, and A. Y. Zomaya, "Some observations on optimal frequency selection in DVFS-based energy consumption minimization," *J. Parallel Distrib. Comput.,* vol. 71, pp. 1154-1164, 2011.

[55] D. Yang, K. Malkowski, P. Raghavan, and M. Kandemir, "Towards energy efficient scaling of scientific codes," in *Parallel and Distributed Processing, 2008. IPDPS 2008. IEEE International Symposium on*, 2008, pp. 1-8.

[56] A. Molnos and K. Goossens, "Conservative Dynamic Energy Management for Real-Time Dataflow Applications Mapped on Multiple Processors," presented at the 12th Euromicro Conference on Digital System Design, Architectures, Methods and Tools, 2009.

[57] Y. Hotta, M. Sato, H. Kimura, S. Matsuoka, T. Boku, and D. Takahashi, "Profile-based optimization of power performance by using dynamic voltage scaling on a pc cluster," presented at the IEEE International Symposium on Parallel and Distributed Processing (IPDPS), 2006.

[58] K. H. Kim, R. Buyya, and J. Kim, "Power Aware Scheduling of Bag-of-Tasks Applications with Deadline Constraints on DVS-enabled Clusters," presented at the Proceedings of the Seventh IEEE International Symposium on Cluster Computing and the Grid, 2007.

[59] R. Springer, D. K. Lowenthal, B. Rountree, and V. W. Freeh, "Minimizing execution time in MPI programs on an energy-constrained, power-scalable cluster," presented at the Proceedings of the eleventh ACM SIGPLAN symposium on Principles and practice of parallel programming, New York, New York, USA, 2006.

[60] B. Rountree, D. K. Lowenthal, S. Funk, V. W. Freeh, B. R. de Supinski, and M. Schulz, "Bounding energy consumption in large-scale MPI programs," in *Supercomputing, 2007. SC '07. Proceedings of the 2007 ACM/IEEE Conference on*, 2007, pp. 1-9.

[61] Q. Qiu and M. Pedram, "Dynamic power management based on continuous-time Markov decision processes," presented at the Proceedings of the 36th annual ACM/IEEE Design Automation Conference, New Orleans, Louisiana, United States, 1999.

[62] T. Simunic, L. Benini, A. Acquaviva, P. Glynn, and G. D. Micheli, "Dynamic voltage scaling and power management for portable systems," presented at the Proceedings of the 38th annual Design Automation Conference, Las Vegas, Nevada, United States, 2001.

[63] S. Irani, S. Shukla, and R. Gupta, "Algorithms for power savings," *ACM Trans. Algorithms,* vol. 3, p. 41, 2007.